\documentclass{epl}
\usepackage{bm}
 
\newcommand{\pd}[2]{\frac{\partial #1}{\partial {#2}}}  
\newcommand{\rref}[1]{(\ref{#1})}
\newcommand{\bq}{\bm {q}}
\newcommand{\bv}{\bm {v}}
\newcommand{\bF}{\bm {F}}
\newcommand{\cH}{{\cal H}}
\newcommand{\cP}{{\cal P}}  
\shorttitle{Nonlinear absorption of SAW by composite fermions} 
\shortauthor{J. Bergli \etal}
  
\title{Nonlinear absorption of surface acoustic waves by composite
fermions
}  
\author{J. Bergli\inst{1},\thanks{E-mail
\email{joakim.bergli@fys.uio.no}}  \And  Y. M.  Galperin\inst{1,2}} 
\institute{
\inst{1}Department of Physics, University of Oslo, PO Box 1048 Blindern, 
 N-0316 Oslo, Norway\\
\inst{2}Solid State Division, A. F. Ioffe   
Physico-Technical  Institute - 194021 St. Petersburg, Russia
and Centre for Advanced Studies -  Drammensveien 78,  
0271 Oslo, Norway
}


\begin{document}
\maketitle  

\begin{abstract} 
Absorption of surface acoustic waves by a two-dimensional electron gas
in a perpendicular magnetic field is considered. The structure  of such
system at the filling factor $\nu$ close to 1/2 can be understood as a
gas of {\em composite fermions}. It is shown that the absorption at
$\nu =1/2 $ can be strongly nonlinear, while small deviation form 1/2
will restore the linear absorption. Study of nonlinear absorption
allows one to determine the force acting upon the composite fermions
from the acoustic wave at turning points of their trajectories.      
\end{abstract}  
  
\section{Introduction}  
The interaction with surface acoustic waves (SAW) is an important tool 
in the study of two-dimensional electron
gases (2DEG) in various
regimes~\cite{Wixforth},
in particular, under conditions of the fractional quantum Hall
effect~\cite{Willett,Simon1996}. 
As well known, the two-dimensional electron system 
exhibits a metallic phase in strong magnetic fields, near the half filled 
Landau level, $\nu=1/2$. This phase has been understood as a gas of so-called
{\em composite Fermions} (CFs). This concept formulated in the
framework of a Chern-Simons theory~\cite{Halperin1993} has appeared
successful to explain 
the acoustic properties of
2DEG~\cite{Simon1998,Mirlin1997b,Willett1997} and to
extract quantitative information about CFs' trajectories. 

So far, both experimental and theoretical studies have concentrated on 
the linear response regime, which is valid for low sound intensities.
On the other hand, a very peculiar {\em nonlinear} response of 3DEG to
acoustic waves (AW) has been predicted~\cite{Galperin1972}
and experimentally observed in \chem{InSb}~\cite{Ivanov} and in
\chem{Ga}~\cite{Fil}. A striking feature of 
this nonlinear response is its {\em anomalous sensitivity} to an external
magnetic field~\cite{Galperin1976}. In fact, nonlinear response
appears suppressed by so weak magnetic field which does not effect
linear absorption at all~\cite{Fil}. The theory of nonlinear response
in external magnetic field has been elaborated
in~\cite{Kozub1975,Galperin1978}, a comprehensive review is given
in~\cite{Galperin1979}. What is important is that an external magnetic
field provides an {\em intrinsic scale} to measure the force acting
upon electrons from the AW.       

It is therefore natural to investigate whether the nonlinear effects
observed for electrons are also present in the two dimensional CF         
liquid, and what new information they provide. This is the aim of the
present paper. 



\section{Background and qualitative discussion}
\begin{figure}[h]
\onefigure[height=5cm]{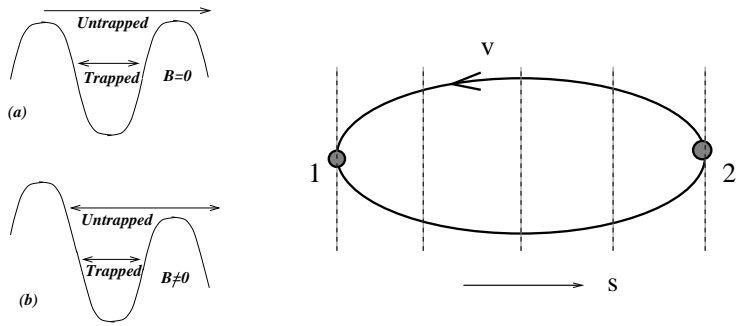}
\caption{Left panel -- On the trapping of resonant electrons in the
absence (a) and in the presence of an external magnetic field
$B$. Right papenl -- Electron trajectories in the magnetic
field. Straight lines indicate the planes of equal phase of the
AW.} \label{fig1} 
\end{figure}

To begin with, let us recall the qualitative picture of nonlinear
acoustic response of 3DEG. If wave vector $\bq$ is much greater than
the electron mean free path $\ell$, then an electron traverses many
acoustic periods before being scattered. Consequently, it contributes
to the absorption as a free particle. Since for typical electron the
$\bq$-projection of the electron velocity $\bv$, $v_q
\equiv (\bq \cdot \bv)/v$, is much greater than the sound velocity, $v_s$,
it ``feels'' a rapidly oscillating field of the acoustic wave, the
contribution to the absorption being small. As  result, only a small
electron group with $v_q \approx v_s$ appears important. These {\em
resonant} electrons determine linear absorption. The situation is very
similar to the well known {\em Landau damping} of plasma waves~\cite{Landau}. 
Turning to the {\em nonlinear} effects one has to discuss dynamics of
the resonant electrons in the finite-amplitude field produced by the
AW. As a result, a part of resonant electrons with small $|v_q-v_s|$
become trapped by the acoustic field. Moving fully synchronously with
the AW, the trapped electrons do not contribute to the absorption, the
total absorption being decreased. This is is the reason for the
nonlinear behavior of the absorption in the absence of the external
magnetic field. The situation is illustrated in 
Fig.~\ref{fig1}(a). 

Now let us turn to the linear absorption in the presence of the
magnetic field. The most interesting behavior takes place at
\begin{equation} 
q^{-1} \ll r_c \ll \ell 
\end{equation} 
where $r_c=v/\omega_c$ is the classical cyclotron radius while $\omega_c$
is the cyclotron frequency. The the electron orbit embeds many
acoustic wave lengths, as it is shown in the right panel of
Fig.~\ref{fig1}. Again, only vicinities of the turning points 1 and 2
are important because only in these regions $v_q$ can be of the order
of $v_s$. Correlation of acoustic phases at the turning points lead to
well known {\em geometric oscillations}~\cite{Bommel,Gurevich}. For
composite fermions geometric oscillations have also been clearly
observed~\cite{Willett}.

To preserve the picture of nonlinear absorption discussed above, the
force component along $\bq$, $ev_yB$,  should be much less than the typical
force from the acoustic wave, $F_0=q\Pi_0$. Here $\Pi_0$ is the amplitude of
the AW-induced potential profile. Otherwise some electrons appear {\em
de-trapped}, and the absorption returns to the linear one. The
situation near a turning point in this case is illustrated in
Fig.~\ref{fig1}(b). Comparing the forces we arrive at the estimate for
the critical magnetic field, $B_c=F_0/v_F$. Measuring $B_c$ one
immediately finds the force $F_0$ near the turning points. To
reproduce the above discussed scheme for CFs is the main idea behind
this work.
       

Now let us turn to the case of interests, namely to 2DEG interacting with SAW.
In many experimental situations one can neglect the deformational
interaction due to strains created by
SAW in the plane of 2DEG, and the force experienced by the electrons
is purely electromagnetic. It is created by the 
external magnetic field and by the AC  electric field of the surface acoustic 
wave. The latter arises because of piezoelectric effect in the
materials used to create the  2DEG (GaAs-AlGaAs heterostructures).
The ``bare'' piezoelectric field is parallel to the propagation
direction, $\bq \parallel \bm {\hat{x}}$. Here $\bm {\hat{x}}$ is the unit
vector along $\bq$-direction. 
The effective electric field is then given by   
$\bm {E} (x, t) = \bm {E}_0 \sin\xi = -\nabla \Phi$ with   
$\Phi = \Phi_0  \cos\xi$. 
Here $\Omega$ is the SAW frequency, and $\xi=qx-\omega t$ is the wave
coordinate;  
$ \bm {E}_0 \parallel\bq \parallel \bm{  
\hat{x}}$. By $\Phi$ we mean the screened electrostatic potential. 
The relationship between this and the bare potential is discussed in 
Ref.~\cite{Simon1996}. The CF-picture arises after the 
Chern-Simons transformation which attaches an even number of flux quanta 
of a fictitious magnetic field to each electron. The resulting particles
are called the {\em composite fermions}. For definiteness we will 
consider the case of two attached flux quanta, appropriate for 
the state around $\nu=1/2$. At the mean field level, the composite 
fermions will then feel an effective magnetic field 
$B^*=B+b$, where $B$ is the external (real) magnetic field and
$b=-2\phi_0n$ is the Chern-Simons field. Here $\phi_0=2\pi\hbar/e$ is 
the magnetic flux quantum and $n$ is the electron density which
includes the density modulation by SAW, $n=n_0+\delta n$. Correspondingly 
there will be a modulation of the Chern-Simons field, $b=b_0+b^{\ab{ac}}$.
We will let $B^*=B+b_0$ represent the average effective field, so that the 
total magnetic field acting on the composite fermions is $B^*+b^{\ab{ac}}$.

In addition, the motion of the   
CFs will create an AC electric Chern-Simons field which is given by  
$\bm {e}^{\ab{ac}}= (2\phi_0/e)\,[\bm {\hat{z}}\times\bm {j}]$.   
The $y$-component of $\bm {e}^{\ab{ac}}$ is given by the $x$-component
of the current. We can find this from the density modulation using 
charge conservation. Assuming that the density modulation is  $\delta
n=(\delta n)_0\cos\xi$, we get  
$j_x=ev_s\delta n$. We will later see that this assumption is justified. 
We then have 
$ e_y^{\ab{ac}}=2\phi_0v_s\delta n $.
This is true under 
the assumption of a harmonic density perturbation and as long as there 
is no net current through the sample. 

The $x$-component of the CS electric field can, 
as explained in Ref.~\cite{Bergli2000}, can be considered as 
a 
potential field.  
The corresponding potential will be denoted $\Psi$, and in the 
following we will calculate the response to this field.
However, it can be shown (see below) that in the regime of strong
nonlinearity the CS electric field is not important at all.
%
%

\section{Theory}\label{Boltzmannsec}
Below we employ the random phase approximation (RPA).
To calculate the nonlinear absorption by CFs we employ the Boltzmann
equation for the CF distribution function, $f$, considering CFs as
particles with charge $-e$ and mass $m$. Consequently, the 
classical  Hamiltonian is 
\begin{equation}\label{resham}
 \cH=(\bm {P}+e\bm {A})^2/2m-e\Psi,\label{Ham}
\end{equation}
where ${\bf P}$ is the canonical momentum (the kinematic momentum 
is $\bm {p}=\bm {P}+e{\bf A}$), ${\bf A}$ is the vector potential
while $\Psi$ is the total scalar potential as explained above.
The vector potential consists of two parts. One emerges from the
static external  
effective magnetic field $B^*$, and one from the AC Chern-Simons field 
that is created by the SAW-induced density modulation.
At $\nu=1/2$, $B^*=0$, and 
the magnetic field is then 
$b^{\ab{ac}}=2\phi_0 \left[n_0 -(2\pi\hbar)^{-2}\int \upd^2P\, f \right]$.

It is convenient to split the distribution function as
$f=f_0(\cH)+f_1$ where $f_0$ is the Fermi function. Then
the Boltzmann equation for $f_1$ is 
\begin{equation}\label{Boltzmann}
\partial f_1/\partial t+\nabla_{\bm{P}}\cH\nabla_{\bm{
 r}}f_1-\nabla_{\bm {r}}\cH\nabla_{\bm {P}}f_1 
 +f_1/\tau = -(\partial \cH/\partial t)(\partial f_0/\partial
 \cH).
\end{equation}
Here we use the relaxation time approximation $-f_1/\tau$ for the
collision operator which significantly simplifies the calculations.
As is emphasized in~\cite{Mirlin1997b}, this leads to charge 
non-conservation. However, this is not expected to give any 
qualitative change at $q\ell \gg 1$ (see, e.~g., \cite{Simon1998}). 
It should be noted that the Hamiltonian~(\ref{Ham}) is written in
terms of the  
AC Chern-Simons magnetic field $b^{\ab{ac}}$. The latter must be
expressed through  
the density modulation as an integral over the distribution function. 
The Boltzmann equation (\ref{Boltzmann}) is then in reality a complicated 
integro-differential equation for the non-equilibrium distribution function. 
It is easy to show, however, that the main contribution to the density
modulation  
comes from the equilibrium part $f_0(\cH)$, so that in calculating
$f_1$ we can approximate the density modulation  
with $\delta n^{(0)}$ coming from $f_0(\cH)$.
Indeed, using 
the fact that in all the region of acoustic amplitudes $e\Psi \ll
\epsilon_F$, where  $\epsilon_F$ is the Fermi energy, 
 we can then expand $f_0(\cH)$ around the point $\cH=p^2/2m$.
The lowest-order term, $\delta n^{(0)}$, is estimated as 
\begin{equation}\label{deltan0}
 \delta n^{(0)} = -e\Psi(2\pi\hbar)^{-2} \int \upd^2p\, 
   (\partial f_0/\partial \cH)\left|_{\cH=p^2/2m} \right.
  = g e\Psi
\end{equation}
Here $g=m/2\pi\hbar^2$ is the density of states per spin (as
usual, we assume the 2DEG to be fully spin-polarized).
Then we can solve Eq.~(\ref{Boltzmann}) for $f_1$ with the assumption 
that $\delta n=\delta n^{(0)}$, and come back to
show that the non-equilibrium correction coming from $f_1$ is small
compared to  $\delta n^{(0)}$.


We will first consider the case $B^*=0$.
Proceeding to the solution, we note that
in the resonant region $v_y\approx v_F$ and $v_x\approx s \ll v_F$. 
The magnetic force
then points mainly in the $x$-direction, and the main part of this
will be given by $b^{ac}v_F$. This may be combined with the potential
$\Psi$ to give the effective potential $\Pi$ such that $-\nabla \Pi=(\pm
v_Fb^{\ab{ac}}+E)\bm {\hat{x}}$, the sign being $+$ for particles with $v_y>0$ and $-$ for $v_y<0$.
Using the approximation $\delta n\approx \delta n^{(0)}$ and assuming the 
electric potential to have the form $\Psi=\Psi_0\cos\xi$ we can find the 
explicit expression for $\Pi$, $\Pi(\xi)=\Pi_0 \psi(\xi)$, where
\begin{equation}
\psi(\xi)=\cos(\xi\mp \theta)\, , \quad  \Pi_0 = \Psi_0
 \sqrt{1+\alpha^2}\, , \quad
 \alpha=2mv_F/q\hbar\, , \quad  \theta \equiv \arctan \alpha \, . 
\end{equation}
The equation for $f_1$ can then be written as 
\begin{equation}\label{resbolt}
 s(\partial f_1/\partial\xi)+\psi'(\partial f_1/\partial s)+af_1=aU,
\end{equation}
with 
\begin{equation}
s=  (v_x-v_s)/\bar{v}, \ \bar{v}^2=e\Pi_0/m\, , \quad 
  U=-\tau e\omega(\partial \Pi/\partial \xi)(\partial f_0/\partial
\cH), \quad a=(q\bar{v}\tau)^{-1}\, .
\end{equation}
The dimensionless parameter $a$ has a clear physical meaning. Indeed,
$\bar v$ is just a typical velocity of the particles trapped in the
potential $\Pi (\xi)$, while $\omega_0 \equiv q{\bar v}$ is their
typical oscillation frequency. Since each scattering event rotates the
particle momentum and leads to its escape from the resonant group,
nonlinear behavior exists only if $\omega_0 \tau \gg 1$, or $a \ll 1$
Thus $a$ is the main parameter responsible for nonlinear behavior.
 Equation (\ref{resbolt}) is easily solved by the method of characteristics, 
giving the equations
\begin{equation}
 \upd \xi/s=\upd s/\psi'(\xi)=\upd f_1/a(U-f_1)\, .
\end{equation}
Solving first the equation for $s$ and $\xi$ we obtain
\begin{equation}
s^2=2(\psi+\eta), \label{eta}
\end{equation}
where $\eta $ is a constant of integration. It has the meaning of a
dimensionless energy for the motion in the $x$-direction.
The remaining equation for $f_1$ and $\xi$ is then
\begin{equation} \label{resbolt1}
 s(\upd f_1/\upd \xi)\pm a f_1=\pm aU\, .
\end{equation}
The sign is $+$ for particles with $s>0$ 
 and $-$ for particles with $s<0$.
Equation (\ref{resbolt1}) requires boundary conditions. For the
 untrapped particles, we use 
 periodic boundary conditions, while for  
the trapped ones we require
\begin{equation}
 f_{1,t}^+(\xi_1)= f_{1,t}^-(\xi_1), \qquad f_{1,t}^+(\xi_2)= f_{1,t}^-(\xi_2),
\end{equation}
where $\xi_1$ and $\xi_2$ are the turning points to the left and right of 
$\xi$ respectively. The result is 
\begin{eqnarray}
 \bar{f}_{1,ut}^+(\xi) 
  &=& \left[e^{a\int_{0}^{2\pi}\frac{\upd \xi''}{s(\xi'')}}-1\right]^{-1}
  \int_{\xi}^{\xi+2\pi}\upd \xi'\frac{\psi'(\xi')}{s(\xi')}
  e^{a\int_{\xi}^{\xi'}\frac{\upd \xi''}{s(\xi'')}}\, ; \label{ut}\\
 \bar{f}_{1,t}^+(\xi) &=&
 \left[\sinh\left\{a\int_{\xi_1}^{\xi_2}\frac{\upd \xi''} 
 {s(\xi'')} \right\}\right]^{-1}
   \int_{\xi_1}^{\xi_2}\upd \xi'\frac{\psi'(\xi')}{s(\xi')}
     \cosh\left\{a\int_{\xi_1}^{\xi'}\frac{\upd \xi''}{s(\xi'')}\right\}
  e^{a\int_{\xi}^{\xi_2}\frac{\upd \xi''}{s(\xi'')}} \nonumber \\
  &&- \int_{\xi}^{\xi_2}\upd \xi'\frac{\psi'(\xi')}{s(\xi')}
  e^{a\int_{\xi}^{\xi'}\frac{\upd \xi''}{s(\xi'')}}, \label{t}
\end{eqnarray}
where we have defined 
\[
 f_1(\xi)= -e\Pi_0\pd{f_0}{H}\frac{\omega}{\omega_0}\bar{f}_1(\xi),
 \qquad \omega_0=q\bar{v}.
\]
The expression for  $\bar{f}_{1,t}^-$ can be obtained by changing the
sign of $s$. 
It is easy to check explicitly, that as $s \gg a$, $f_1 \propto
s^{-1}$. Consequently, only resonant particles are important.

In the following we will consider the limiting case of strong nonlinearity
$a\ll 1 $. In this limit we can expand the distribution functions
(\ref{ut}) and (\ref{t}) in powers of $a$ and then compute the
non-equilibrium density modulation $\delta n^{(1)}$. As a result,
$\delta n^{(1)}/\delta n^{(0)} \approx a \alpha v_s/v_F$ 
which is small ($v_s/v_F$ is of order $1/30$ in 
typical experiments. The approximation could break down if $\alpha$ 
was very large).
Similarly, we may calculate the current in the $y$-direction due to
$f_1$ in order to  
determine the $x$-component of the CS electric field. Again, this
contribution appears proportional to $a$ and small.
 This means that 
any higher harmonics in the CS electric field will be suppressed 
by a factor $a$. 

\section{Nonlinear absorption}\label{absorption}
We are now able to calculate the absorbed power per length of cross
section, $P$,  from the acoustic 
wave. It is given by 
\begin{equation}
 P=\int\frac{\upd^2p}{(2\pi\hbar)^2}\langle\dot \cH f\rangle
= \left(e\Pi_0/2\pi\right)^2(v_s/v_F)\, g \omega \cP \, ,
\end{equation}
where $\langle\cdots\rangle$ denotes average over the period of the acoustic 
wave, while
\begin{equation}\label{pb}
 \cP=-\sum_{\pm}\int_0^{2\pi}\upd \xi\int \upd \eta\,
 \frac{\psi'(\xi)}{|s(\xi)|}\bar{f}_1^{\pm}(\xi,\eta). \label{cP}
\end{equation}
When evaluating $\cP$ one must include contributions from both trapped
and untrapped particles in all directions (i. e. for both signs
$\pm$), and adjust the range of $\eta$ 
accordingly. The contributions from particles
with different sign of $s$ will cancel in the order $a^0$ terms, 
and the leading contribution will be to order $a$.
After rather tedious calculations we arrive at the result,
\begin{equation}
 P=C(e\Pi_0/2\pi)^2 \, (v_s/v_F)\, g \omega a \, ,
\end{equation}
where $C \sim 1$ is some numerical factor that is found from numerical
integration.

\section{Effect of a weak effective magnetic field}\label{magnetic}
Let us now turn to the case where the effective external magnetic 
field, $B^*$, is not exactly zero, that is, where the filling fraction 
$\nu$ is close to but not equal to $1/2$. 
In the resonant region, the Hamiltonian  \rref{resham} is then changed to 
\begin{equation}
 \cH'=\cH_{\ab{res}}+ev_FB^*x.
\end{equation}
Here and in the following we write the expressions 
for particles with $v_y>0$, for $v_y<0$ the sign of 
the last term must be changed.
We still write $f=f_0(\cH_{\ab{res}})+f_1$, but the force acquires a new contribution 
$\bF=e\nabla \Pi-ev_FB^*$. The Boltzmann equation \rref{resbolt} is then transformed
to the form 
\begin{equation}
 s(\partial f_1/\partial \xi)+[b-\psi'(\xi)](\partial
 f_1/\partial s)+af_1=aU, \qquad b=v_FB^*/q\Pi_0\, .
\end{equation} 
The physical interpretation of the parameter $b$ is the ratio of the magnetic 
force from the external magnetic field to the force from the modified 
electrostatic potential $\Pi (\xi)$. Solving the equations for the
characteristics we get instead of Eq.~(\ref{eta}), 
\begin{equation}
 s^2/2=\psi(\xi) -b\xi+\eta.
\end{equation}
Here we write $\xi$ for $\xi-\theta$, i. e. we translate the origin of the 
coordinates to adjust to the modified potential. In the case $b\gg1$, 
in which we are most interested, there will be only one turning 
point, which we will denote $\xi_0$, and all particles will be untrapped.
This will then be the only point where
$s=0$, and we get $\eta = b\xi_0-\psi(\xi_0)$. The particles
will then come from  
$\xi=-\infty$ at $t=-\infty$, turn at $\xi_0$ and return to 
$\xi=-\infty$ at $t=\infty$. Of course this is not the true trajectory 
of the particles, but it approximates the true trajectory close to the 
turning point. The nonequilibrium distribution function is 
\begin{equation}
 f_1(\xi,\eta)=a\int_{-\infty}^{\xi}d\xi'\,
  \frac{U}{s(\xi,\eta)}e^{-a\int_{\xi'}^{\xi} 
  \frac{d\xi''}{s}}.
\end{equation}
Here the integral over $\xi$ is to be taken along the trajectory
defined by the constant $\eta$. That is, for 
particles that have passed their turning point (and thus have $s<0$) we 
must integrate up to $\xi_0$ and then back to $\xi$ remembering the 
change of sign of $s$. When $a\ll1$ and $b\gg1$ the argument of the 
exponential is very close to 0 in the region of effective interaction 
near the turning point. Following Ref.~\cite{Galperin1976}
we will therefore set the exponential to 1 in this 
region. Changing variable from the integration constant $\eta$ to the
position of the turning point, $\xi_0$, and expanding $1/s$ in powers
of $1/b$  to lowest order we obtain for the total absorption 
\begin{equation}\label{abssb}
 P=(\pi/2)\, (e\Pi_0/2\pi)^2 (v_s/v_F)g \omega \, . 
\end{equation}
This is equal to the {\em linear absorption}, $P_0$, 
in the absence of an external magnetic field, which can be directly
calculated from the linearized Boltzmann equation.
 Indeed, magnetic field restores linear 
absorption. 
\section{Discussion} The possibility to restore linear absorption at
\begin{equation} \label{bstar}
B^*=B-B_{1/2} \ge B_c \equiv q\Pi_0/v_F\, , \quad B_{1/2} \equiv
B\left|_{\nu=1/2} \right. \, ,  
\end{equation}  
allows one to determine directly the $\bq$-component of the force,
$qe\Pi_0$, acting 
upon CF at the turning points. Since the effective force is a
complicated function of SAW intensity and frequency (a detailed theoretical
study of this force will be published elsewhere), the relationship
(\ref{bstar}) seems useful. On the other hand, the product $q\Pi_0$
can be directly determined from Eq.~\rref{abssb} knowing the   {\em measured}
absorbed power, $P$. Consequently, a way to check the above concept is
first to reach nonlinear behavior at $B=B_{1/2}$, then restore the
linear behavior by changing magnetic field by the quantity $\ge B_c$,
and finally measure the absorbed power without changing SAW intensity.  

\section{Conclusion}
It is shown that absorption of SAW by 2DEG will show a pronounced
nonlinear behavior at $\nu=1/2$. Small deviations from $\nu =1/2$ will
restore linear absorption. Studies of these deviations allow one to
determine the effective force acting upon composite fermions.   

\section{Acknowledgements} Part of this work has been done during the visit of the authors to the Weizmann
Institute of Science, Rehovot, Israel.
 Support from the OEC
Project -- \textit{Access to Submicron Center for Research on
Semiconductor Materials, Devices and Structures} (HPRI-CT-1999-0026)
is acknowledged.

\end{document}